\documentclass[aps,prb,twocolumn,showpacs,superscriptaddress,groupedaddress]{revtex4} 

\usepackage[dvipdfm]{graphicx} 
\usepackage{subfigure}
\usepackage{dcolumn} 
\usepackage{amsmath}
\usepackage{txfonts}
\usepackage{bm} 

\begin{document}

\widetext

\title{Superconductivity assisted by interlayer pair hopping in multilayered cuprates} 

\author{Kazutaka Nishiguchi} 
\affiliation{Department of Physics, The University of Tokyo, Hongo, Tokyo 113-0033, Japan} 

\author{Kazuhiko Kuroki} 
\affiliation{Department of Physics, Osaka University, 1-1 Machikaneyama, Toyonaka, Osaka 560-0043, Japan} 

\author{Ryotaro Arita} 
\affiliation{Department of Applied Physics, The University of Tokyo, Hongo, Tokyo 113-8656, Japan} 
\affiliation{JST, PRESTO, Kawaguchi, Saitama 332-0012, Japan} 

\author{Takashi Oka} 
\affiliation{Department of Physics, The University of Tokyo, Hongo, Tokyo 113-0033, Japan} 

\author{Hideo Aoki} 
\affiliation{Department of Physics, The University of Tokyo, Hongo, Tokyo 113-0033, Japan}

\date{\today}

\begin{abstract} 
In order to explore why the multilayered cuprates have such high $T_{\text{c}}$'s, 
we have examined various interlayer processes. 
Since the interlayer one-electron hopping has little effects on the band structure, 
we turn to the interlayer pair hopping. 
The superconductivity in a double-layer Hubbard model with and without the interlayer pair hopping, 
as studied by solving the Eliashberg equation with the fluctuation exchange approximation, 
reveals that the interlayer pair hopping acts to increase the pairing interaction and the self-energy simultaneously, 
but that the former effect supersedes the latter and enhances the superconductivity, 
along with how the sign of the interlayer off-site pair hopping determines the relative configuration of $d$-waves between the adjacent layers. 
Study of the triple-layer case with the interlayer pair hopping further reveals that 
the superconductivity is further enhanced but tends to be saturated toward the triple-layer case. 
\end{abstract}

\maketitle

\section{Introduction} 
Although we are witnessing the discovery of new classes of superconductors that include the iron-based and organic superconductors,\cite{Scalapino12, Uemura09, Aoki12} 
the high-$T_{\text{c}}$ cuprate superconductors stand out in having the highest-$T_{\text{c}}$ to date. 
Specifically, among various families of the cuprate, 
the highest $T_{\text{c}}$ occurs in the multilayered cuprates that have $n$ CuO$_{2}$ planes in a unit cell, 
typically the Hg-based HgBa$_2$Ca$_{n-1}$Cu$_n$O$_{2n+2+\delta}$ (Hg-$12(n-1)n$), 
where $T_{\text{c}}$ depends on the number, $n$, of the CuO$_{2}$ planes with $T_{\text{c}}$ increasing for $n=1$ to $3$ and decreasing slightly for $n \geq 4$, 
and Hg-$1223$ is still the highest $T_{\text{c}}$ superconductor.\cite{Schilling93} 
Empirically, the electronic band structure has been probed with ARPES for the Bi-based triple-layer cuprate (Bi-$2223$).\cite{Ideta10} 
Another experiment examines the optical Josephson plasma modes\cite{Kleiner92} arising from interlayer Josephson couplings 
from the reflectivity spectra in the Hg-based multilayered cuprates for $n=2$ to $5$, 
where the change in the Josephson coupling strength is shown to be correlated with $T_{\text{c}}$.\cite{Hirata12} 

There have been several theoretical studies for multilayered cuprates: 
Anderson and Chakravarty proposed that 
an interlayer Josephson coupling that arises as a process second-order in the interlayer one-electron hopping enhances the superconductivity.\cite{Anderson97b, Chakravarty93} 
Although this mechanism may be related to the $c$-axis coherence, 
it is considered to be insufficient for increasing $T_{\text{c}}$ 
because the realistic magnitude of $t_{z}$ is an order of magnitude smaller than the intralayer one ($t$), 
so the interlayer Josephson pair tunneling ($\propto t_z^2/t$) in this 
picture is too small to enhance the superconductivity. 
Chakravarty also studied the effect of the interlayer Josephson pair tunneling phenomenologically 
in a macroscopic Ginzburg-Landau free energy scheme.\cite{Chakravarty04} 
On the other hand, Leggett examined a Coulomb energy in the $c$-axis layering structure,\cite{Leggett99} 
while Okamoto {\it et al.} studied an effect of the interlayer one-electron hopping for double-layer Hubbard and $t$-$J$ models.\cite{Okamoto08} 
Chen {\it et al.} have also examined an effect of the interlayer tunneling in terms of a free energy derived from $t$-$J$ model 
in a case where the phenomenological interlayer coupling is chosen to realize the in-phase gap function between the two layers.\cite{Chen12} 
Given the background, our purpose here is to {\it microscopically} investigate 
a mechanism of the superconductivity in multilayered cuprates 
focusing on the effects of {\it microscopic interlayer pair hopping}. 
We envisage that the interlayer pair hopping arises as the matrix elements of long-range Coulomb interaction, 
rather than a process second-order in the interlayer one-electron hopping or phenomenological Josephson coupling. 

Motivated by this, 
we start from a double-layer Hubbard model 
to explore microscopically the multilayered cuprates by examining various interlayer processes. 
The interlayer one-electron hopping has turned out to exert little effects on the first-principles band structure (not shown), 
so that we turn to the interlayer pair hopping. 
The hopping of Cooper pairs across the layers should in general exist as a matrix element of the long-range Coulomb interaction,\cite{Kusakabe12, Kusakabe09} 
and this should affect superconductivity as a process intrinsic in multilayer systems, 
but whether and how the superconductivity is enhanced has not been well understood. 
Since we are talking about $d$-wave pairing that basically mediated by antiferromagnetic spin fluctuations around specific regions in $k$-space, 
we have to adopt a method that can incorporate $k$-dependent pairing interactions. 
Hence we adopt here fluctuation exchange (FLEX) approximation,\cite{KadanoffBaym61, Baym62, BSW89, Bickers89, Dahm95} 
whose result is fed into the Eliashberg equation. 
We shall show that 
the interlayer pair hopping acts both ways to increase the pairing interaction and decrease the quasi-particle life time (with an increased self-energy), 
but the former effect is found to supersedes the latter and enhances superconductivity, 
along with how the sign of the interlayer off-site pair hopping determines the relative configuration of $d$-wave between the adjacent layers. 
We have extended the study to the triple-layer case with the interlayer pair hopping, 
where we show that the superconductivity is further enhanced but only sublinearly with the number of layers with a tendency for saturation toward the triple-layer case. 

\section{Formalism} 
We consider a Hamiltonian of the double-layer model $H$ with the interlayer pair hopping $H_{\text{pair}}$, 
\begin{equation} 
H= H_{t} +H_{U} +H_{\text{pair}},
\end{equation} 
where the one-electron kinetic energy, 
\begin{equation} 
H_{t}= \sum_{\alpha \beta} \sum_{ij} \sum_{\sigma} t^{\alpha \beta}_{ij} c^{\alpha \dagger}_{i \sigma} c^{\beta}_{j \sigma}, 
\end{equation} 
and the Hubbard interaction, 
\begin{equation} 
H_{U}= U \sum_{\alpha} \sum_{i} c^{\alpha \dagger}_{i \uparrow} c^{\alpha \dagger}_{i \downarrow} c^{\alpha}_{i \downarrow} c^{\alpha}_{i \uparrow}, 
\end{equation} 
are defined in a usual way, with $c^{\alpha \dagger}_{i \sigma}$ creating an electron at $i$-th site with spin $\sigma$ in the layer $\alpha$, 
$ t^{\alpha \beta}_{ij} $ the transfer integral and $U$ the on-site Coulomb repulsion.   
$H_{t}$ consists of the intralayer ($\alpha=\beta$) 
and interlayer ($\alpha\neq \beta$) one-electron hoppings, 
where the intralayer component is here considered for the nearest-neighbor $t=-0.5$ eV, 
second-neighbor $t^{\prime}=0.1$ eV up to the third-neighbor $t^{\prime \prime}=-0.08$ eV. 
For the interlayer one-electron hopping we take a usually adopted form, 
\begin{equation} 
H_{t\perp}= \sum_{\alpha \neq \beta} \sum_{\bm{k}} \sum_{\sigma} 
            \frac{t_{z}}{2} \left( \cos k_{x} -\cos k_{y} \right)^2 c^{\alpha \dagger}_{\bm{k} \sigma} c^{\beta}_{\bm{k} \sigma}, 
\end{equation} 
in $k$-space,\cite{Andersen94, Andersen95} with $t_{z}=-0.05$ eV here. 
These values of the one-electron hoppings are basically determined by a downfolding from the first-principles bands, 
but we here make a simplification in which we take common values between the single-, double-, and triple-layer cases for a transparent comparison. 

\begin{figure}[htbp]
\begin{center}
\subfigure[]{ \includegraphics[clip, width=6.0cm]{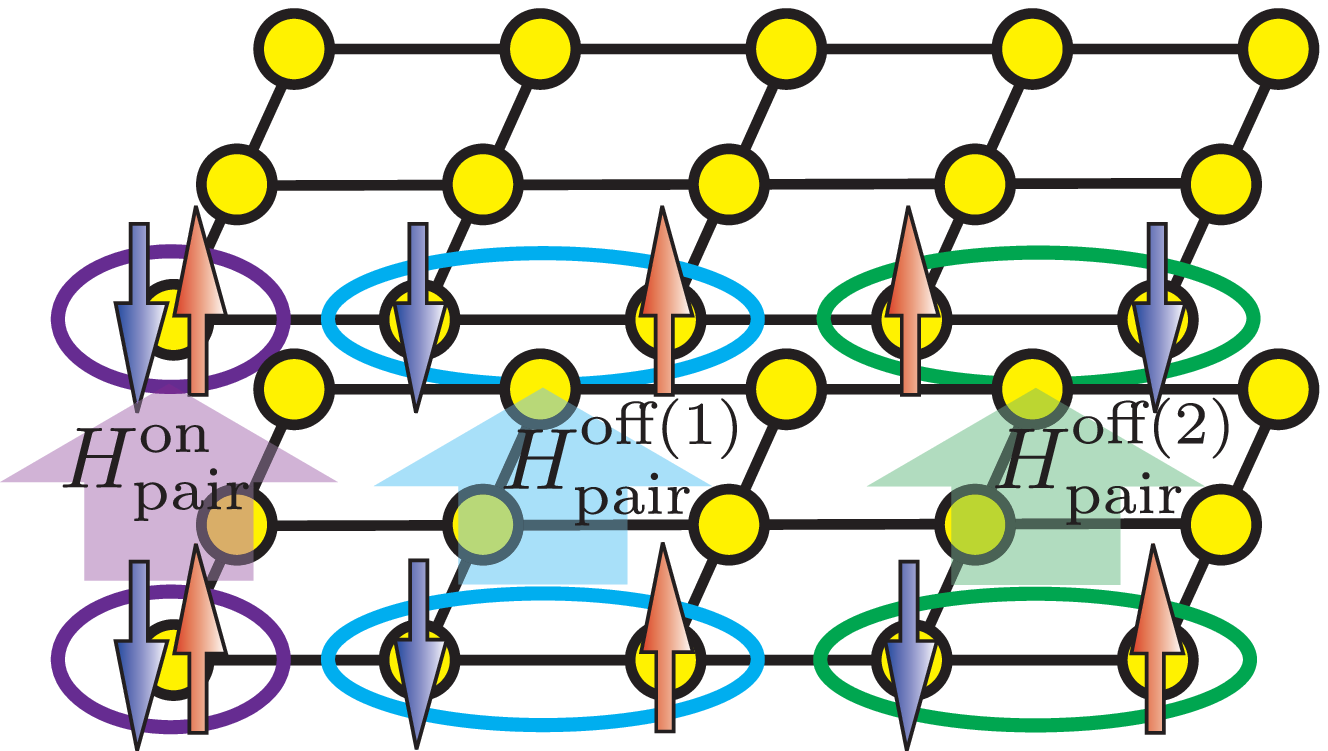} } 
\subfigure[]{ \includegraphics[clip, width=6.0cm]{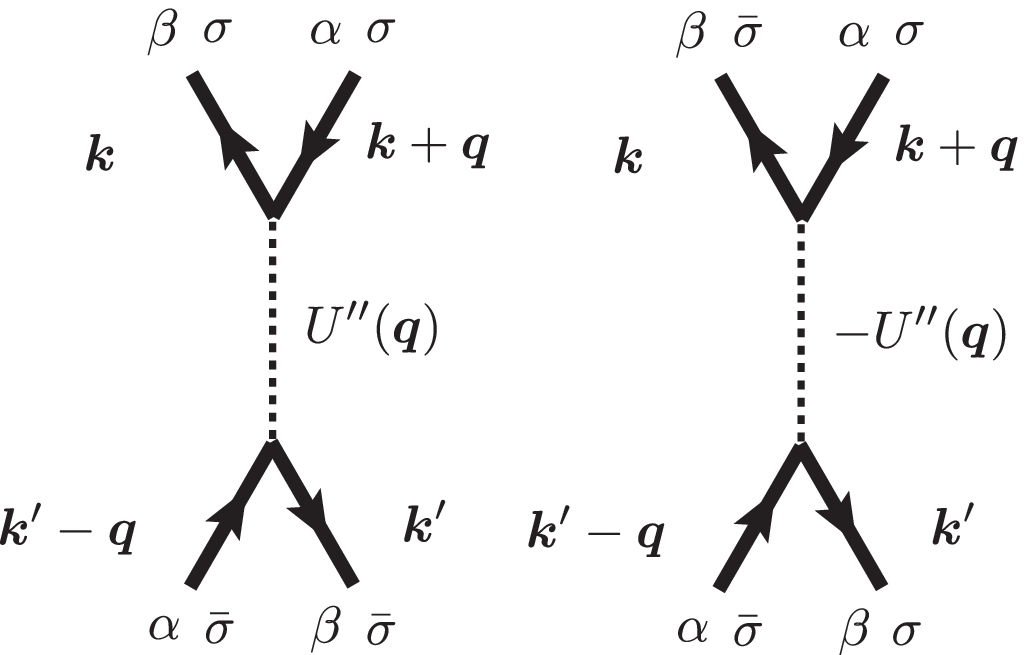} \label{fig:Vnsfsf} } 
\caption{(color online). 
(a) Schematic interlayer hopping of on-site pairs ($H^{\text{on}}_{\text{pair}}$ ) and off-site pairs ($H^{\text{off}}_{\text{pair}}$). 
The latter consists of $H^{\text{off}(1)}_{\text{pair}}$ and $H^{\text{off}(2)}_{\text{pair}}$, 
where the spins of the pair are exchanged during the hop in $H^{\text{off}(2)}_{\text{pair}}$. 
(b) Diagrams for the non-spin-flip interaction $H^{\text{off}(1)}_{\text{pair}}$ 
(left panel) and spin-flip interaction $H^{\text{off}(2)}_{\text{pair}}$ (right). } 
\label{fig:PHfig} 
\end{center} 
\end{figure}

Now the question is the form of 
the interlayer pair hopping $H_{\text{pair}}$.  
Here we take a rather general form 
$H_{\text{pair}}= H^{\text{on}}_{\text{pair}} +H^{\text{off}}_{\text{pair}}$, 
where in addition to the usually considered interlayer on-site pair hopping,  
\begin{gather} 
H^{\text{on}}_{\text{pair}}= U^{\prime} \sum_{\alpha \neq \beta} \sum_{i} 
c^{\alpha \dagger}_{i \uparrow} c^{\alpha \dagger}_{i \downarrow} c^{\beta}_{i \downarrow} c^{\beta}_{i \uparrow},
\end{gather} 
we also consider {\it interlayer off-site pair hopping}, 
$H^{\text{off}}_{\text{pair}}= H^{\text{off}(1)}_{\text{pair}} +H^{\text{off}(2)}_{\text{pair}}$, 
where the first term, 
\begin{gather}  
H^{\text{off}(1)}_{\text{pair}}= U^{\prime \prime} \sum_{\alpha \neq \beta} \sum_{ij}^{\text{nn}} 
c^{\alpha \dagger}_{i \uparrow} c^{\alpha \dagger}_{j \downarrow} c^{\beta}_{j \downarrow} c^{\beta}_{i \uparrow}, 
\end{gather}
is the hopping of a spin-singlet pair formed on nearest-neighbor intralayer sites from one layer to another, 
with $\sum_{ij}^{\text{nn}}$ denoting a sum over nearest-neighbors. 
In addition, we have to note that, 
if we want to preserve the spin SU(2) symmetry, 
we should include 
\begin{gather}  
H^{\text{off}(2)}_{\text{pair}}= U^{\prime \prime} \sum_{\alpha \neq \beta} \sum_{ij}^{\text{nn}} 
c^{\alpha \dagger}_{i \uparrow} c^{\alpha \dagger}_{j \downarrow} c^{\beta}_{i \downarrow} c^{\beta}_{j \uparrow}, 
\end{gather} 
in which the spins of the pair are exchanged during the hop (FIG. \ref{fig:PHfig}). 
While the on-site term is considered to be the largest interlayer pair hopping, 
the off-site terms should be not only the second largest interlayer pair hopping arising from long-range Coulomb interaction, 
but may also play a crucial role for $d$-wave pairing. 

Now, the FLEX approximation, 
which is a conserved approximation with bubble and ladder diagrams included\cite{KadanoffBaym61, Baym62}, 
is one of the standard methods for self-consistently treating the spin- and charge-fluctuation mediated pairing 
with the self-energy effect incorporated\cite{BSW89, Bickers89, Dahm95}.  
Let us start with showing that 
the method can be extended for treating the pair-hopping processes introduced here.  
Derived from Dyson-Gor'kov equation, 
the linearized Eliashberg equation for the gap function $\Delta_{\alpha \beta}(k)$ reads in the present case, 
\begin{equation} 
\begin{split}
\lambda \Delta_{\alpha \beta}(k) 
&= -\frac{1}{N\beta} \sum_{k^{\prime}} \sum_{\alpha^{\prime} \beta^{\prime}} \sum_{\gamma \delta} 
    V^{\text{pair}}_{\alpha^{\prime} \alpha \beta \beta^{\prime}}(k-k^{\prime})  \\ 
& \times G_{\alpha^{\prime} \gamma}(k^{\prime}) \Delta_{\gamma \delta}(k^{\prime}) G_{\beta^{\prime} \delta}(-k^{\prime}). 
\label{eq:gapeq}
\end{split} 
\end{equation} 
Here $k=(\bm{k}, \omega_{n})$ is the two-dimensional wave number and Matsubara frequency for fermions with a $32 \times 32 \times 2048$ mesh, 
$\beta= 1/T$ ($k_{B}=1$), and $\lambda$ the eigenvalue of the Eliashberg equation, 
where $T_\text{c}$ is identified from $\lambda=1$ 
but $\lambda$ also serves as a measure of the strength of superconductivity. 
The pairing interaction $\hat{V}^{\text{pair}}$, which is equivalent to the effective interaction $\hat{V}^{F}$ for the anomalous Green's function, 
being involved with layer index in the present case, 
becomes a bit complicated (a $2 \times 2 \times 2 \times 2$ tensor) as 
\begin{equation} 
V^{\text{pair}}_{\alpha^{\prime} \alpha \beta \beta^{\prime}}(q)= 
V^{F}_{\alpha^{\prime} \alpha \beta \beta^{\prime}}(q) \equiv 
\Bigg[ \hat{U} +\frac{3}{2} \frac{\hat{U} \hat{\chi}_{0} \hat{U}}{1-\hat{U} \hat{\chi}_{0}} 
               -\frac{1}{2} \frac{\hat{U} \hat{\chi}_{0} \hat{U}}{1+\hat{U} \hat{\chi}_{0}} 
\Bigg]_{\alpha^{\prime} \alpha \beta \beta^{\prime}}(q) , 
\label{eq:Veff} 
\end{equation} 
where 
$[\hat{\chi}_{0}]_{\alpha \alpha^{\prime} \beta \beta^{\prime}}(q) = 
-(1/N\beta) \sum_{k} G_{\beta \alpha}(k+q) G_{\alpha^{\prime} \beta^{\prime}}(k)$ 
is the polarization function, 
while $\hat{U}$, also a $2 \times 2 \times 2 \times 2$ tensor, represents the interaction, 
which can be expressed as a $4 \times 4$ matrix, 
with the four rows (columns) corresponding to 
$\alpha \alpha^{\prime} (\beta \beta^{\prime}) = 11, 22, 12, 21$, as 
\begin{equation}
\hat{U}(q)= 
\begin{pmatrix} 
U & 0 & 0 & 0 \\
0 & U & 0 & 0 \\ 
0 & 0 & 0 & U^{\prime} +U^{\prime \prime}(\bm{q}) \\ 
0 & 0 & U^{\prime} +U^{\prime \prime}(\bm{q}) & 0 
\end{pmatrix} 
\end{equation} 
with $U^{\prime \prime}(\bm{q})= 2U^{\prime \prime} \left( \cos q_{x} +\cos q_{y} \right)$. 

For each layer, the $d$-wave pairing is favored by the 
intralayer pairing interaction $V^{\text{pair}}_{\alpha \alpha \alpha \alpha}(q)$ 
that has peaks around $\bm{Q}= (\pm \pi, \pm \pi)$\cite{BSW89, Bickers89, Dahm95, KK.HA97}. 
For the multilayered model with the interlayer pair hopping, 
the question is how the interlayer pairing interaction $V^{\text{pair}}_{\alpha \beta \beta \alpha}(q)$ $(\alpha \neq \beta)$ affects superconductivity. 

\section{Results} 
\subsection{Eigenvalues of the Eliashberg equation} 
\begin{figure}[htbp]
\begin{center}
\includegraphics[width=8.0cm,clip]{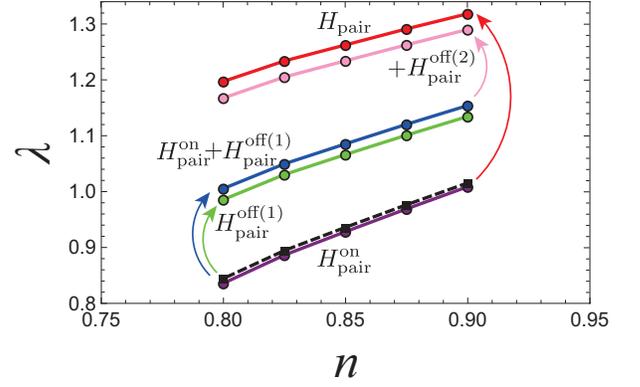} 
\caption{(color online). 
Eigenvalue $\lambda$ of 
Eliashberg equation against the band 
filling $n$ for the double-layer model at $T=0.01$ eV. 
Black (dashed) line: no interlayer pair hopping, 
purple: with $H^{\text{on}}_{\text{pair}}$ only, 
green: with $H^{\text{off}(1)}_{\text{pair}}$ only, 
blue: with $H^{\text{on}}_{\text{pair}} +H^{\text{off}(1)}_{\text{pair}}$, 
pink: with $H^{\text{on}}_{\text{pair}} +H^{\text{off}(1)}_{\text{pair}}$ and isolated diagrams for $H^{\text{off}(1)}_{\text{pair}}$, 
red: with all of $H_{\text{pair}}$ except for mixing of $H^{\text{off}(1)}_{\text{pair}}$ and $H^{\text{off}(2)}_{\text{pair}}$. 
} 
\label{fig:EigVal}
\end{center} 
\end{figure}

Now we present the results comparing the situations in the presence and absence of the interlayer pair hopping in FIG. \ref{fig:EigVal}. 
This plots the eigenvalues of the Eliashberg equation $\lambda$ against the band filling $n$, 
where we set $U=2.5$ eV here to be a relatively small value compared to the realistic parameter 
but appropriate to FLEX which is a weak-coupling formalism. 
For the interlayer pair hopping, we set $U^{\prime}=-2U^{\prime \prime}=0.5$ eV to be much smaller than $U$ but still significant, 
while the effect of the sign $U^{\prime \prime}$ will be discussed later. 

\begin{figure}[htbp]
\begin{center}
\includegraphics[width=8.cm,clip]{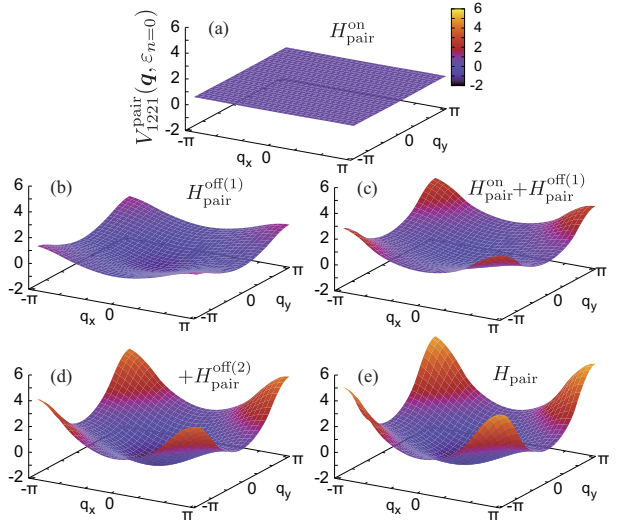} 
\caption{(color online). 
Real part of $V^{\text{pair}}_{1221}(\bm{q}, \varepsilon_{n=0})$ 
when various interlayer pair hoppings are switched on one by one at $n=0.85$. 
(a) $H^{\text{on}}_{\text{pair}}$ only, 
(b) $H^{\text{off}(1)}_{\text{pair}}$ only, 
(c) $H^{\text{on}}_{\text{pair}} +H^{\text{off}(1)}_{\text{pair}}$, 
(d) $H^{\text{on}}_{\text{pair}} +H^{\text{off}(1)}_{\text{pair}}$ and isolated diagrams for $H^{\text{off}(2)}_{\text{pair}}$, 
(e) All of $H_{\text{pair}}$ except for mixing of $H^{\text{off}(1)}_{\text{pair}}$ and $H^{\text{off}(2)}_{\text{pair}}$.} 
\label{fig:Vip} 
\end{center}
\end{figure}

Beside the eigenvalues of the Eliashberg equation $\lambda$, 
we also display in FIG. \ref{fig:Vip} the interlayer pairing interaction $V^{\text{pair}}_{1221}(q)$ $( =V^{\text{pair}}_{2112}(q) )$ (at $n=0.85$). 
This is important since the $d$-wave pairing within each layer has a strongly $k$-dependent form, 
$\Delta_{11}(\bm{k})= \Delta_{22}(\bm{k}) \sim \cos k_{x} -\cos k_{y}$, 
so that the real question for multilayered cases should be the effect of interlayer pair hoppings on such anisotropic gap functions. 

In order to resolve the effects from various terms, let us switch on the terms one by one. 
First, black dashed line in FIG. \ref{fig:EigVal} represents the result of the double-layer model without interlayer pair hopping. 
Since FLEX becomes unreliable when the band filling becomes too close to the half-filling, 
we only plot the result up to $n \lesssim 0.9$. 
Now, purple line in FIG. \ref{fig:EigVal} is for the model with interlayer on-site pair hopping $H^{\text{on}}_{\text{pair}}$ only, 
where the pair hopping is seen to {\it suppress} the superconductivity in fact. 
This result, 
which may at first seem strange since an interlayer pairing interaction would naively enhance the superconductivity, 
comes from the following fact. 
The interlayer pair hopping does produce an interlayer pairing interaction as displayed in Fig. \ref{fig:Vip}, 
which is expected to enhance the intralayer superconducting gap functions $\Delta_{11}(\bm{k})$ and $\Delta_{22}(\bm{k})$ 
in the sense of Suhl-Kondo mechanism\cite{Suhl59, Kondo63}. 
However, the interlayer pair hopping also increases the (intralayer) self-energy. 
An increased self-energy is a bad news for superconductivity, 
and the result here indicates that this effect supersedes the enhanced interlayer pairing interaction. 
If we look at FIG. \ref{fig:Vip}(a), 
the interlayer pairing interaction $V^{\text{pair}}_{1221}(q)$ only shows barely visible peaks around $\bm{Q}$. 
This is because the interlayer on-site pair hopping Hamiltonian has no $k$-dependence to start with, 
and FLEX diagrams do not render a significant $k$-dependence. 
This is why $V^{\text{pair}}_{1221}(q)$ is insufficient for overcoming the increased self-energy. 

By sharp contrast, 
green line in FIG. \ref{fig:EigVal}, 
which represents the result when one of interlayer off-site pair hopping $H^{\text{off}(1)}_{\text{pair}}$ is switched on (without $H^{\text{on}}_{\text{pair}}$), 
exhibits a significant enhancement. 
Indeed, FIG. \ref{fig:Vip}(b) shows that $V^{\text{pair}}_{1221}(q)$ develops a significant $k$-dependence, 
and this is how the enhanced pairing interaction overcomes the increased self-energy, 
since $H^{\text{off}(1)}_{\text{pair}} (\propto \cos q_{x} +\cos q_{y}$) originally possesses a large $k$-dependence, 
where the peaks around $\bm{Q}$ are intensified in FLEX scheme. 

Now the question is whether the addition of the on-site pair hopping ($H^{\text{on}}_{\text{pair}}$) degrades the enhancement due to $H^{\text{off}(1)}_{\text{pair}}$.  
Blue line in FIG. \ref{fig:EigVal} representing this situation shows that 
the superconductivity is enhanced {\it even above} the case when $H^{\text{off}(1)}_{\text{pair}}$ alone is switched on. 
This may first seem to contradict with the fact that $H^{\text{on}}_{\text{pair}}$ suppresses the superconductivity, 
but, if we go to FIG. \ref{fig:Vip}(c), 
$V^{\text{pair}}_{1221}(q)$ with both of $H^{\text{on}}_{\text{pair}}$ and $H^{\text{off}(1)}_{\text{pair}}$ switched on 
is more reinforced around $\bm{Q}$ than when $H^{\text{off}}_{\text{pair}}$ alone is present. 
The increase in $V^{\text{pair}}_{1221}(q)$ is caused by the process 
in which $H^{\text{on}}_{\text{pair}}$ raises the peaks of $V^{\text{pair}}_{1221}(q)$ around $\bm{Q}$ from $H^{\text{off}(1)}_{\text{pair}}$ 
through the spin-fluctuation term $\frac{3}{2} \frac{\hat{U} \hat{\chi}_{0} \hat{U}}{1-\hat{U} \hat{\chi}_{0}}$ in (\ref{eq:Veff}). 

Let us now discuss the effects of the other interlayer off-site pair hopping $H^{\text{off}(2)}_{\text{pair}}$. 
The term is required for the SU(2), but, being a spin-flip interaction, does complicate the diagrams as follows. 
The term reads in $k$-space as 
$H^{\text{off}(2)}_{\text{pair}}= 
(-1/N)\sum_{\alpha \neq \beta} \sum_{\bm{k} \bm{k}^{\prime} \bm{q}} U^{\prime \prime}(\bm{q}) 
c^{\alpha \dagger}_{\bm{k}+\bm{q} \downarrow} 
c^{\alpha \dagger}_{\bm{k}^{\prime} +\bm{q} \uparrow} 
c^{\beta \dagger}_{\bm{k}^{\prime} \downarrow} 
c^{\beta \dagger}_{\bm{k} \uparrow} $ 
as depicted in FIG. \ref{fig:PHfig}(b). 
We can readily extend the FLEX when all the interactions are of the spin-flip form, 
where the formulation is similar to the usual FLEX. 

Therefore we first take account of $H^{\text{on}}_{\text{pair}} +H^{\text{off}(1)}_{\text{pair}}$, 
and the isolated diagrams for $H^{\text{off}(2)}_{\text{pair}}$ separately 
(i.e., excluding the mixing of $H^{\text{off}(2)}_{\text{pair}}$ with non-spin-flip $H^{\text{on}}_{\text{pair}} +H^{\text{off}(1)}_{\text{pair}}$). 
In this case, the pairing interaction $\hat{V}^{\text{eff}}$ in Eq. (\ref{eq:Veff}) is replaced with 
$\hat{V}^{F}[U,U^{\prime},U^{\prime \prime}] +\hat{V}^{F}[0,0,U^{\prime \prime}]$. 
For details, see the appendix below. 
Dramatically, the addition of $H^{\text{off}(2)}_{\text{pair}}$ is seen as pink line in FIG. \ref{fig:EigVal} to enhance the superconductivity 
much more than the case with $H^{\text{on}}_{\text{pair}}$ and $H^{\text{off}(1)}_{\text{pair}}$ alone. 
A reinforced $V^{\text{pair}}_{1221}(q)$ around $\bm{Q}$ are indeed seen in FIG. \ref{fig:Vip}(d). 

Finaly, we take account of the mixing of $H^{\text{on}}_{\text{pair}} +H^{\text{off}(1)}_{\text{pair}}$ with $H^{\text{off}(2)}_{\text{pair}}$. 
To treat this rigirously is difficult 
because they have respective $\bm{k}$-dependences, and their mixing acts as a kind of vertex corrections (see the appendix). 
However, we have confirmed from the self-energy that the effect of the vertex corrections is numerically negligible, 
so that we can take account of all of $H_{\text{pair}}$ except for the mixing of $H^{\text{off}(1)}_{\text{pair}}$ and $H^{\text{off}(2)}_{\text{pair}}$ 
by replacing $\hat{V}^{\text{eff}}$ in Eq. (\ref{eq:Veff}) with 
$\hat{V}^{F}[U,U^{\prime},U^{\prime \prime}] 
+\hat{V}^{F}[U,U^{\prime},U^{\prime \prime}] 
-\hat{V}^{F}[U,U^{\prime},0]$. 
The first (second) terms represent the non-spin-flip (spin-flip) interactions, 
while the third term subtracts the double counting (see the appendix). 
Red line in FIG. \ref{fig:EigVal} represents the result in this scheme, 
where the superconductivity is enhanced even above the pink one. 
FIG. \ref{fig:Vip}(e) confirms that 
$V^{\text{pair}}_{1221}(q)$ is more reinfored around $\bm{Q}$ than in FIG. \ref{fig:Vip}(d). 
The increase in $V^{\text{pair}}_{1221}(q)$ (pink line to red in FIG. \ref{fig:EigVal}) is caused in FLEX  
because a combined effect of $H^{\text{on}}_{\text{pair}}$ and $H^{\text{off}(2)}_{\text{pair}}$ raises $V^{\text{pair}}_{1221}(q)$, 
as a combined effect of $H^{\text{on}}_{\text{pair}}$ and 
$H^{\text{off}(1)}_{\text{pair}}$ raises $V^{\text{pair}}_{1221}(q)$ (green line to blue).

\subsection{Phase diagram} 
\begin{figure}[htbp]
\begin{center}
\includegraphics[width=8.0cm,clip]{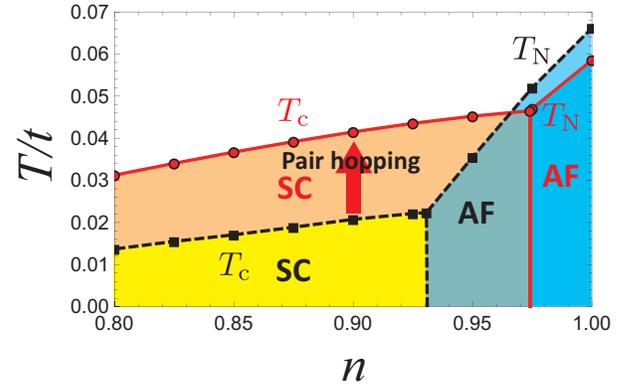} 
\caption{(color online). 
Phase diagram on $T$ and $n$ (carrier concentration) 
for the double-layer system with (red  lines) and without (black) interlayer pair hopping $H_{\text{pair}}$. 
$T_{\text{c}}$ is SC transition temperature while $T_{\text{N}}$ AF transition (Ne\'{e}l) temperature.
The arrow represents the increase of $T_{\text{c}}$ arising from the interlayer pair hopping. } 
\label{fig:PD2}
\end{center}
\end{figure}

Now, we are in position to construct a phase diagram of the double-layer system, 
in which we can compare the result with and without interlayer pair hopping $H_{\text{pair}}$ in FIG. \ref{fig:PD2}.  
Superconducting (SC) phase boundary is identified from the eigenvalue of linearized Eliashberg equation $\lambda$ reaching unity. 
The antiferromagnetic (AF) phase boundary is determined in a usually adopted way from the (in the present case the intralayer) 
$[ \hat{U} \hat{\chi}_{0} ]_{\alpha \alpha \alpha \alpha}$ 
approaching unity ($0.975$ here). 

As is seen in FIG. \ref{fig:PD2}, 
SC transition temperature $T_{\text{c}}$ for double-layer model 
in the presence of the interlayer pair hopping $H_{\text{pair}}$ 
is higher than the case in the absence for all the range of the carrier concentration considered here. 
For $U^{\prime}=-2U^{\prime \prime}=0.5$ eV, 
the increase of $T_{\text{c}}$ amounts to $\Delta T_{\text{c}} \sim 0.02t \sim 100$ K. 
On the other hand, 
AF transition temperature $T_{\text{N}}$ for the double-layer model with interlayer pair hopping $H_{\text{pair}}$ 
slightly decreases from the case without, which 
is because the divergence of the spin susceptibility $\chi^{\text{s}}_{\alpha \alpha \alpha \alpha}$ is suppressed 
by the self-energy increased due to the interlayer pair hopping. 

\subsection{Configuration of the $d$-wave pairing} 
\begin{figure}[htbp] 
\begin{center}
\subfigure[$U^{\prime \prime}<0$.]{ \includegraphics[width=4.0cm,clip]{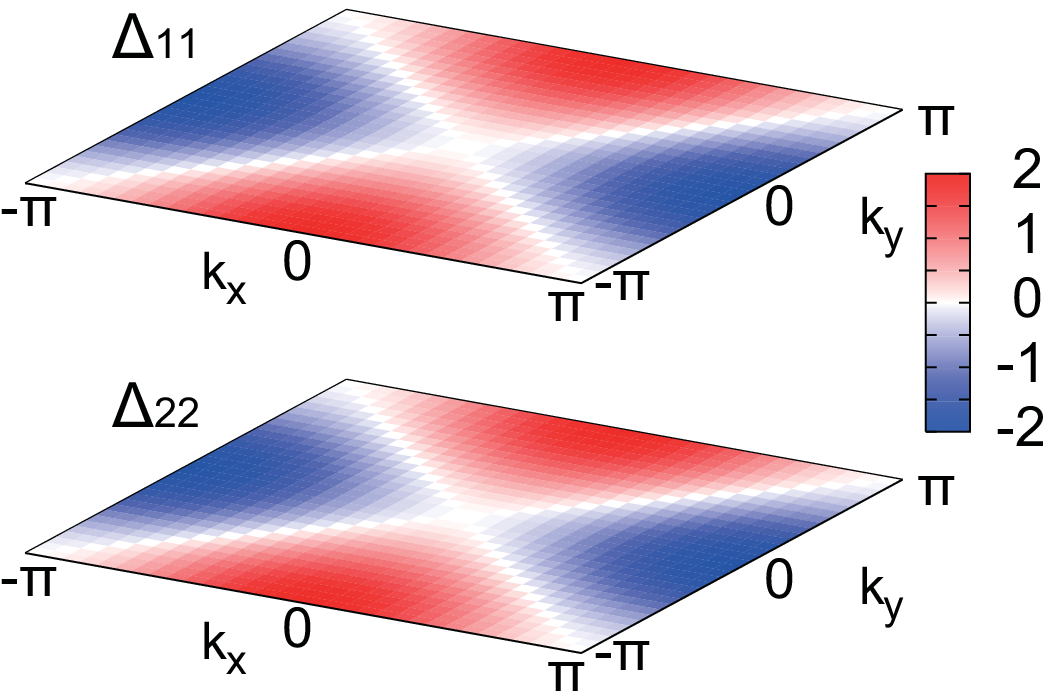} \label{fig:dconfig-0} } 
\subfigure[$U^{\prime \prime}>0$.]{ \includegraphics[width=4.0cm,clip]{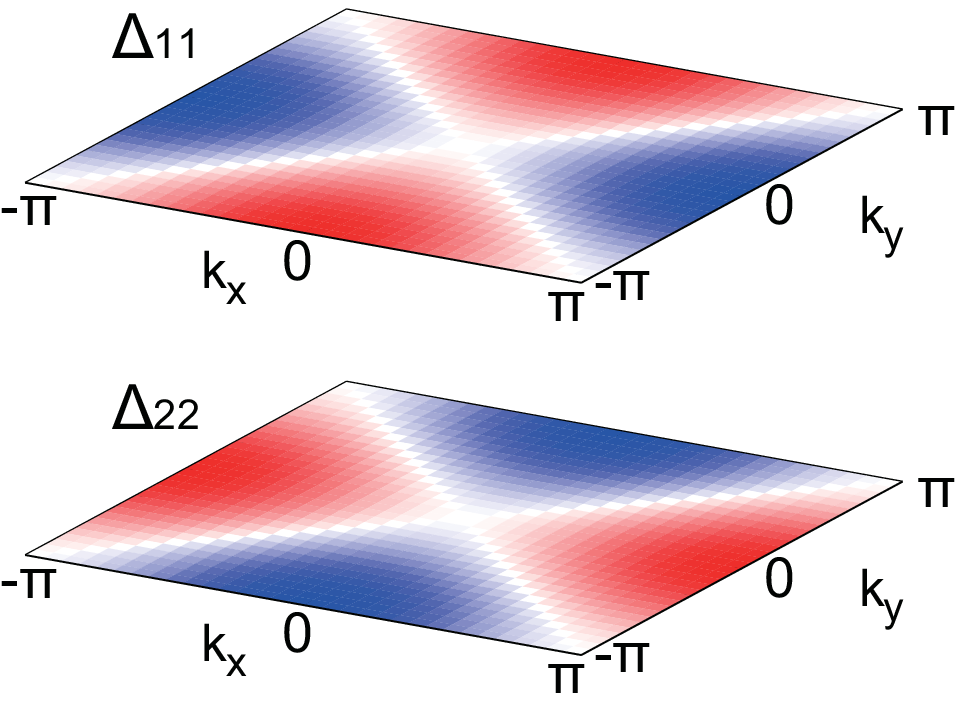} \label{fig:dconfig-pi} } 
\caption{(color online). 
In-plane gap functions for the top layer 
($\Delta_{11}$) and for the bottom layer ($\Delta_{22}$) for 
$U^{\prime \prime}<0$ with an in-phase configuration (a) and 
for $U^{\prime \prime}>0$ with an out-of-phase one (b)}.
\label{fig:d-conf} 
\end{center}
\end{figure}

Now a word on the 
sign of $U^{\prime \prime}$ in $H^{\text{off}}_{\text{pair}}$.  
The interlayer pairing interaction $V^{\text{pair}}_{1221}(q)$ with $U^{\prime \prime}<0$, as we have assumed so far, favors 
the configuration where the in-plane $d$-wave gap functions $\Delta_{11}$ and $\Delta_{22}$ are arrayed in-phase as in FIG. \ref{fig:dconfig-0}.  
If we have $U^{\prime \prime}>0$, on the other hand, 
we end up with a configuration where $\Delta_{11}$ and $\Delta_{22}$ are arrayed 
out-of-phase as in FIG. \ref{fig:dconfig-pi}, where 
$V^{\text{pair}}_{1221}(q)$ also changes sign. 

To be more precise, however, 
the configuration is not determined solely by the sign of $U^{\prime \prime}$: 
even in the absence of the interlayer pair hopping, the in-phase configuration is favored 
through the off-diaginal Green's functions, $G_{12}$ and $G_{21}$, in the Eliashberg equation (\ref{eq:gapeq}). 
When the interlayer pair hopping is switched on, 
the effect of $V^{\text{pair}}_{1221}(q)$ has to overcome this effect of $V^{\text{pair}}_{1111}(q)$ which favors the in-phase configuration 
before out-of-phase configuration is realized for large enough $U^{\prime \prime}>0$.

\subsection{Triple-layer} 
\begin{figure}[htbp]
\begin{center}
\includegraphics[width=7cm,clip]{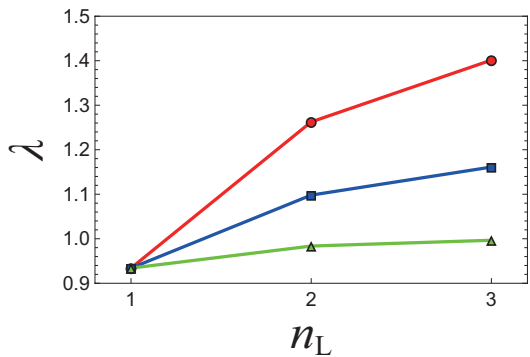} 
\caption{(color online). 
Eigenvalue $\lambda$ of the 
Eliashberg equation against the number of layers $n_{\text{L}}$ 
for the average band filling $n=0.85$ at $T=0.01$ eV. 
We vary $U^{\prime}=-2U^{\prime \prime}=0.5$ eV (red line), 
$0.3$ eV (blue), $0.1$ eV (green). 
} 
\label{fig:123Up} 
\end{center}
\end{figure} 
Finally, we discuss the effect of interlayer pair hopping in a triple layer system. 
We include all terms in $H_{\text{pair}}$ except for the mixing, i.e., $H^{\text{off}(1)}_{\text{pair}}$ and $H^{\text{off}(2)}_{\text{pair}}$, for clarity, 
and we assume that the interlayer pair hopping takes place only between the adjacent layers. 
The result, displayed in FIG. \ref{fig:123Up}, shows that 
the eigenvalue $\lambda$ of the Eliashberg equation plotted against the number of layers $n_{\text{L}}$ at the average band filling $n=0.85$ indicates that 
the superconductivity is enhanced monotonically for $n_{\text{L}}=1 \rightarrow 2$ and $2 \rightarrow 3$. 
However, the increase is only sublinear for $n_{\text{L}}$. 
The tendency of saturation is for all the values of $U^{\prime}=-2U^{\prime \prime}$, varied here over $0.1$ - $0.5$ eV. 
We have saturation because, 
although superconductivity in the inner plane (IP) is assisted by interlayer pairing interaction between two outer planes (OPs), 
the self-energy effect becomes stronger since IP interacts with two OPs. 
Therefore the net effect makes the enhancement sublinear. 
This is supported by the following observation: 
The \textit{d}-wave superconducting gap function in IP $\Delta_{22}$ is relatively larger than that in OPs $\Delta_{11}$ and $\Delta_{33}$, i.e., $\Delta_{22} > \Delta_{11}=\Delta_{33}$, 
while the self-energy in IP $\Sigma_{22}$ is also relatively larger than that in OPs $\Sigma_{11}$ and $\Sigma_{33}$, i.e., $\Sigma_{22} > \Sigma_{11}=\Sigma_{33}$.  

\begin{figure}[htbp]
\begin{center} 
\includegraphics[width=4.4cm,clip]{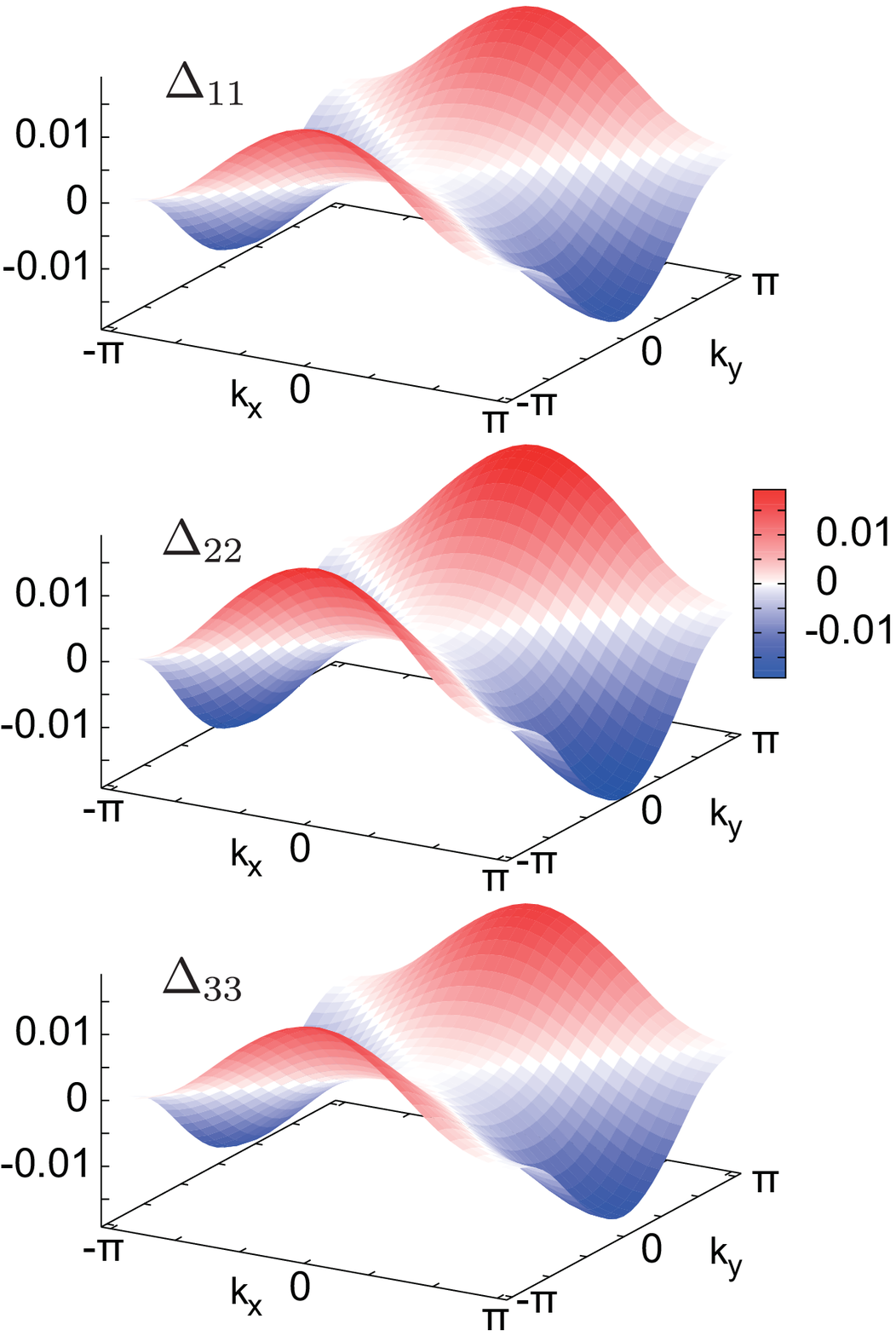} 
\includegraphics[width=4.0cm,clip]{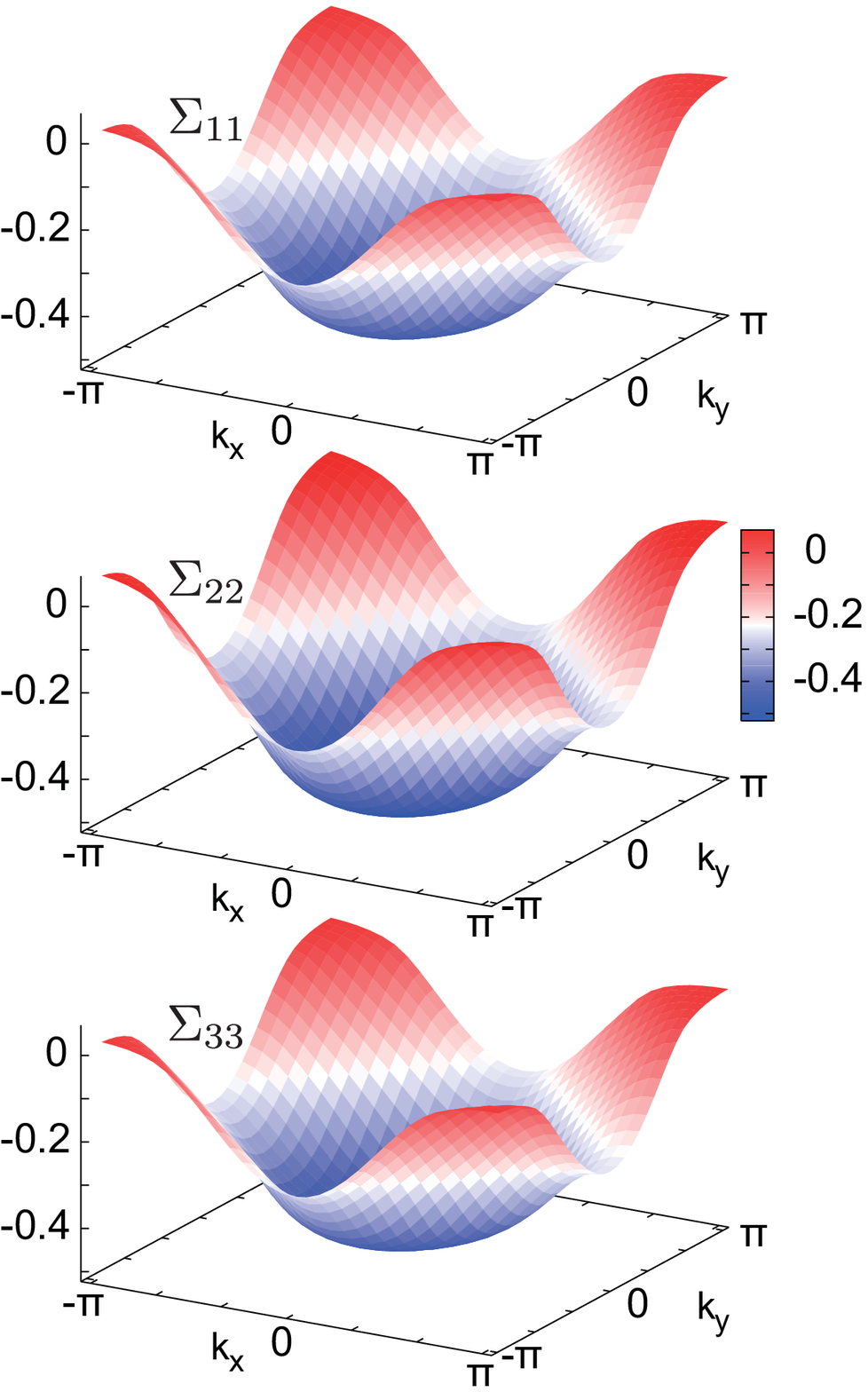} 
\caption{
(Left) Superconducting gap functions in IP and OPs for the triple-layer model with the interlayer pair hopping. 
The \textit{d}-wave gap in IP $\Delta_{22}$ is relatively larger than that in OPs $\Delta_{11}$ and $\Delta_{33}$, i.e., $\Delta_{22} > \Delta_{11}=\Delta_{33}$. 
(Right) Self-energy in IP and OP for the triple-layer model with the interlayer pair hopping. 
The self-energy in IP $\Sigma_{22}$ is relatively larger than that in OPs $\Sigma_{11}$ and $\Sigma_{33}$, i.e., $\Sigma_{22} > \Sigma_{11}=\Sigma_{33}$. 
} 
\label{fig:Gap3l}
\end{center}
\end{figure}

\section{Summary and Discussion} 
To summarize, 
superconductivity in a double-layer Hubbard model with and without the interlayer pair hopping is studied 
by solving the Eliashberg equation with the fluctuation exchange approximation. 
We have shown that 
the interlayer pair hopping acts to increase both the pairing interaction and the self-energy, 
but that 
the former effect supersedes and the latter enhances the superconductivity. 
The interlayer pair hopping considered here is for off-site pairs, 
for which we have found that the extra off-site pair-hopping term needed to preserve SU(2) symmetry, 
actually acts to enhance the superconductivity even further. 
The off-site interlayer pair hopping especially acts to enhance the superconductivity even further. 
We then end up with a phase diagram for the double-layer model 
where the superconducting boundary is significantly 
higher than the case without interlayer pair hopping. 
We also investigate the triple-layer model with the interlayer pair hopping, 
where the superconductivity is further enhanced 
but the enhancement is sublinear for $n_{\text{L}}=1 \rightarrow 3$. 

In evaluating the present mechanism, 
an estimate (e.g., with constrained random phase approximation\cite{Aryasetiawan04} (c-RPA)) of the magnitude of interlayer off-site pair hopping $H^{\text{off}}_{\text{pair}}$ 
in real materials should be important. 
Experimentally, 
one possibly relevant quantity is the optical Josephson plasma energy, 
which has been observed for Hg-based cuprates with 2-5 layers.\cite{Hirata12} 
It is an interesting future problem to examine the actual relation of this to the interlayer pair hopping considered here. 
Also, larger numbers of layers are interesting, for which the study is under way. 

\section*{ACKNOWLEDGMENTS} 
We wish to thank Koischi Kusakabe, Naoto Tsuji, and Takahiro Morimoto for useful discussions.   
This study has been supported by Grants-in-Aid for Scientific Research from
JSPS (Grants No. 23340095, R.A.; No. 22340093, K.K. and H.A.). 
R.A. acknowledges financial support from JST-PRESTO.

\appendix 

\section{FLEX in the coexistence of non-spin-flip and spin-flip interlayer pair hoppings} 
We present an outline of the extension of the 
FLEX (fluctuation exchange) approximation 
to include the spin-flip as well as non-spin-flip interactions.  

In the present context, 
we start with reformulating the multiorbital FLEX with intra- and interlayer interactions in a double-layer model. 
We first separate non-spin-flip interactions such as $H_{U}$, $H^{\text{on}}_{\text{pair}}$ and $H^{\text{off}(1)}_{\text{pair}}$ 
from spin-flip interactions such as $H^{\text{off}(2)}_{\text{pair}}$. 
The non-spin-flip part ($H_{U}$, $H^{\text{on}}_{\text{pair}}$ and $H^{\text{off}(1)}_{\text{pair}}$) can be expressed as 
\begin{equation} 
H_{\text{nsf}} 
= \frac{1}{N} \sum_{\bm{k} \bm{k}^{\prime} \bm{q}} \sum_{\alpha \beta \beta^{\prime} \alpha^{\prime}} 
U^{\text{nsf}}_{\alpha \alpha^{\prime} \beta^{\prime} \beta}(\bm{q})
c^{\alpha \dagger}_{\bm{k}+\bm{q} \uparrow} 
c^{\beta \dagger}_{\bm{k}^{\prime} -\bm{q} \downarrow} 
c^{\beta^{\prime}}_{\bm{k}^{\prime} \downarrow} 
c^{\alpha^{\prime}}_{\bm{k} \uparrow} ,
\end{equation} 
where $\alpha$, $\beta$, etc denote the layer, $\bm{q}$ the momentum transfer, 
and the nonzero components in the present model are 
$ U^{\text{nsf}}_{1111}(\bm{q})= U^{\text{nsf}}_{2222}(\bm{q}) = U $, 
$ U^{\text{nsf}}_{1221}(\bm{q})= U^{\text{nsf}}_{2112}(\bm{q})=  U^{\prime} +U^{\prime \prime}(\bm{q}) $. 
On the other hand, the spin-flip term $H^{\text{off}(2)}_{\text{pair}} 
= -(1/N) \sum_{\bm{k},\bm{k}^{\prime},\bm{q}} \sum_{\alpha \neq \beta} 
U^{\prime \prime}(\bm{q}) 
c^{\alpha \dagger}_{\bm{k}+\bm{q} \downarrow} 
c^{\alpha \dagger}_{\bm{k}^{\prime} -\bm{q} \uparrow} 
c^{\beta}_{\bm{k}^{\prime} \downarrow} 
c^{\beta}_{\bm{k} \uparrow}$, 
which is required for SU(2) be preserved, can be expressed as 
\begin{equation} 
H_{\text{sf}} 
= -\frac{1}{N} \sum_{\bm{k} \bm{k}^{\prime} \bm{q}} \sum_{\alpha \beta \beta^{\prime} \alpha^{\prime}} 
U^{\text{sf}}_{\alpha \alpha^{\prime} \beta^{\prime} \beta}(\bm{q})
c^{\alpha \dagger}_{\bm{k}+\bm{q} \downarrow} 
c^{\beta \dagger}_{\bm{k}^{\prime} -\bm{q} \uparrow} 
c^{\beta^{\prime}}_{\bm{k}^{\prime} \downarrow} 
c^{\alpha^{\prime}}_{\bm{k} \uparrow}, 
\end{equation} 
where the form $c^{ \dagger}_{\downarrow}c^{ \dagger}_{\uparrow}c_{\downarrow}c_{\uparrow}$ signifies the spin-flip, 
and the nonzero components in the present model are 
$ U^{\text{sf}}_{1221}(\bm{q})= U^{\text{sf}}_{2112}(\bm{q})= U^{\prime \prime}(\bm{q}) $. 
While $H_{U}$ and $H^{\text{on}}_{\text{pair}}$ can also be expressed in a spin-flip form (see below), 
we cannot cast both of $H^{\text{off}(1)}_{\text{pair}}$ and $H^{\text{off}(2)}_{\text{pair}}$ simultaneously 
into a single expression like above 
if we want to have the prefactor as a function of $\bm{q}$. 

\begin{figure}[htbp]
\begin{center}
\includegraphics[width=3cm,clip]{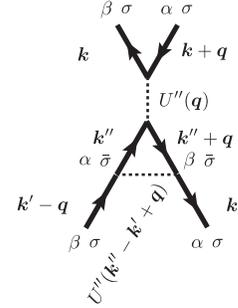} 
\caption{
Second-order cross term between $H^{\text{off}(1)}_{\text{pair}}$ and $H^{\text{off}(2)}_{\text{pair}}$.} 
\label{fig:vrtx}
\end{center} 
\end{figure}

In FLEX scheme, all of the bubble and ladder diagrams composed of $H_{\text{nsf}}$ and $H_{\text{sf}}$ 
have to be summed, which include cross terms of $H^{\text{off}(1)}_{\text{pair}}$ and $H^{\text{off}(2)}_{\text{pair}}$. 
It is difficult to treat the cross terms exactly, 
since a kind of ``vertex correction" as shown in FIG. \ref{fig:vrtx} exists already in the second-order in the perturbation expansion. 
Fortunately, however, we have confirmed numerically that 
such diagrams are much smaller than the other terms in the same order, 
which is because the momentum dependence is different between $H^{\text{off}(1)}_{\text{pair}}$ and $H^{\text{off}(2)}_{\text{pair}}$. 
We can therefore ignore the diagrams composed of the mixing of $H^{\text{off}(1)}_{\text{pair}}$ and $H^{\text{off}(2)}_{\text{pair}}$. 

With this, we can actually sum all the bubble and ladder diagrams for both of $H_{\text{nsf}}$ and $H_{\text{sf}}$ 
(except for the mixing of $H^{\text{off}(1)}_{\text{pair}}$ and $H^{\text{off}(2)}_{\text{pair}}$), 
which is performed as follows. 
The FLEX for $H_{U}$, $H^{\text{on}}_{\text{pair}}$ and $H^{\text{off}(1)}_{\text{pair}}$ 
(i.e., all the bubble and ladder diagrams composed of $H_{\text{nsf}}$) can be performed in a standard way, 
where the only difference is to take into account the tensorial interactions and susceptibilities 
(i.e., $\hat{U}(\bm{q}), \hat{\chi}_{0}(q)$). 

\begin{figure}[htbp]
\begin{center}
\includegraphics[width=4cm,clip]{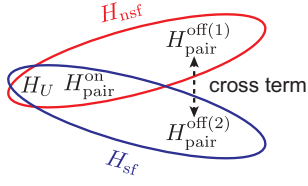} 
\caption{FLEX can be performed for each of the Hamiltonian 
components encircled by ovals.  In addition, cross terms 
exist between the components indicated by an arrow.} 
\label{fig:FLEXfig}
\end{center} 
\end{figure}

We next take account of the mixing of $H_{U}$, $H^{\text{on}}_{\text{pair}}$ with $H^{\text{off}(2)}_{\text{pair}}$ 
employing the following technique. 
First, we cast $H_{U}$ and $H^{\text{on}}_{\text{pair}}$ into a spin-flip form, 
by rearranging creation and annihilation operators, as 
\begin{equation} 
\begin{split} 
H_{U} 
&= -\frac{U}{N} \sum_{\bm{k} \bm{k}^{\prime} \bm{q}} \sum_{\alpha} 
c^{\alpha \dagger}_{\bm{k}+\bm{q} \downarrow} 
c^{\alpha \dagger}_{\bm{k}^{\prime} -\bm{q} \uparrow} 
c^{\alpha}_{\bm{k}^{\prime} \downarrow} 
c^{\alpha}_{\bm{k} \uparrow} , \\ 
H^{\text{on}}_{\text{pair}} 
&= -\frac{U^{\prime}}{N} \sum_{\bm{k} \bm{k}^{\prime} \bm{q}} \sum_{\alpha \neq \beta} 
c^{\alpha \dagger}_{\bm{k}+\bm{q} \downarrow} 
c^{\alpha \dagger}_{\bm{k}^{\prime} -\bm{q} \uparrow} 
c^{\beta}_{\bm{k}^{\prime} \downarrow} 
c^{\beta}_{\bm{k} \uparrow}, 
\end{split} 
\end{equation} 
note that the form $c^{ \dagger}_{\downarrow}c^{ \dagger}_{\uparrow}c_{\downarrow}c_{\uparrow}$ signifies the spin-flip. 
The $H_{U}$ and $H^{\text{on}}_{\text{pair}}+H^{\text{off}(2)}_{\text{pair}}$ 
in the spin-flip form have nonzero components 
$ U^{\text{sf}}_{1111}(\bm{q})= U^{\text{sf}}_{2222}(\bm{q})= U $, 
$ U^{\text{sf}}_{1221}(\bm{q})= U^{\text{sf}}_{2112}(\bm{q})= U^{\prime} +U^{\prime \prime}(\bm{q}) $. 
We can now take account of the mixing of $H_{U}$, $H^{\text{on}}_{\text{pair}}$ and $H^{\text{off}(2)}_{\text{pair}}$ 
when all of the bubble and ladder diagrams composed of $H_{\text{sf}}$ are summed 
(see FIG. \ref{fig:FLEXfig}). 
When we use the technique above, 
the effective interaction for the normal self-energy composed of $H_{\text{nsf}}$ and $H_{\text{sf}}$, 
$\hat{V}^{G\text{.nsf}}[U, U^{\prime}, U^{\prime \prime}(\bm{q})]$ 
and $\hat{V}^{G\text{.sf}}[U, U^{\prime}, U^{\prime \prime}(\bm{q})]$ respectively, 
are equivalent, 
and the pairing interaction for the anomalous self-energy composed of $H_{\text{nsf}}$ and $H_{\text{sf}}$, 
$\hat{V}^{F\text{.nsf}}[U, U^{\prime}, U^{\prime \prime}(\bm{q})]$ 
and $\hat{V}^{F\text{.sf}}[U, U^{\prime}, U^{\prime \prime}(\bm{q})]$ respectively, 
are also equivalent. 

Finally, 
the diagrams composed of $H_{\text{nsf}}$ and those composed of $H_{\text{sf}}$ are added, 
but we have of course to subtract the double-counted diagrams composed of $H_{U}$ and $H^{\text{on}}_{\text{pair}}$. 
This is achieved by putting the effective interaction 
$\hat{V}^{\text{eff}}$ for normal self-energy $\hat{\Sigma}^{G}$ 
and the pairing interaction $\hat{V}^{\text{pair}}$ 
for the anomalous self-energy $\hat{\Sigma}^{F}$ as 
\begin{equation} 
\begin{split} 
\hat{V}^{\text{eff}} 
&= \hat{V}^{G\text{.nsf}}[U, U^{\prime}, U^{\prime \prime}(\bm{q})] 
  +\hat{V}^{G\text{.sf}} [U, U^{\prime}, U^{\prime \prime}(\bm{q})]  \\ 
& \qquad -\hat{V}^{G\text{.(n)sf}}[U, U^{\prime}, 0],  
\label{eq:tVeff} 
\end{split} 
\end{equation} 

\begin{equation} 
\begin{split} 
\hat{V}^{\text{pair}} 
&= \hat{V}^{F\text{.nsf}}[U, U^{\prime}, U^{\prime \prime}(\bm{q})] 
  +\hat{V}^{F\text{.sf}} [U, U^{\prime}, U^{\prime \prime}(\bm{q})]  \\ 
& \qquad -\hat{V}^{F\text{.(n)sf}}[U, U^{\prime}, 0].  
\label{eq:tVpair} 
\end{split} 
\end{equation} 

\subsection{FLEX for Non-spin-flip interactions} 
We first write down the multiorbital FLEX with intra- and interlayer interactions belonging to $H_{\text{nsf}}$. 
The normal self-energy for the interlayer interactions is given as 
\begin{equation}
\Sigma^{G\text{.nsf}}_{\alpha \beta}(k) 
= \frac{1}{N\beta} \sum_{k^{\prime}} \sum_{\alpha^{\prime} \beta^{\prime}} 
  V^{G\text{.nsf}}_{\alpha^{\prime} \alpha \beta^{\prime} \beta}(k-k^{\prime}) 
  G_{\alpha^{\prime} \beta^{\prime}}(k^{\prime}), 
\end{equation} 
where 
\begin{align}
\hat{V}^{G\text{.nsf}} 
&= \hat{U}^{\text{nsf}} +\hat{V}^{G\text{.oB}} +\hat{V}^{G\text{.L}},  \\
V^{G\text{.oB}}_{\alpha^{\prime} \alpha \beta^{\prime} \beta}(q) 
&= \Bigg[ \frac{\hat{U} \hat{\chi}_{0} \hat{U}}
               {1 -\hat{U} \hat{\chi}_{0} \hat{U} \hat{\chi}_{0}} 
   \Bigg]^{\text{nsf}}_{\alpha^{\prime} \alpha \beta^{\prime} \beta}(q) ,\\
V^{G\text{.L}}_{\alpha^{\prime} \alpha \beta^{\prime} \beta}(q) 
&= \Bigg[ \frac{\hat{U} \hat{\chi}_{0} \hat{U} \hat{\chi}_{0} \hat{U}}
               {1 -\hat{U} \hat{\chi}_{0}} 
   \Bigg]^{\text{nsf}}_{\alpha^{\prime} \alpha \beta^{\prime} \beta}(q). 
\end{align}
Here $V^{G\text{.oB}}$ is 
the bubble-diagram contribution to the effective interaction 
for the normal self-energy, 
where odd numbers of bubbles are included due to the spin selection rule in $H_{\text{nsf}}$, 
while $V^{G\text{.L}}$ is 
the ladder-diagram contribution to 
the effective interaction.   
The polarization function is defined as
\begin{equation}
[\hat{\chi}_{0}]_{\alpha \alpha^{\prime} \beta \beta^{\prime}}(q) = 
-\frac{1}{N\beta} \sum_{k} G_{\beta \alpha}(k+q) G_{\alpha^{\prime} \beta^{\prime}}(k), 
\end{equation} 
which is a $2 \times 2 \times 2 \times 2$ tensor for the double-layer model 
and can also be expressed as a $4 \times 4$ matrix. 
As for the products of tensors, we have 
\begin{equation}
[ \hat{U} \hat{\chi}_{0} ]^{\text{nsf}}_{\mu \mu^{\prime} \nu \nu^{\prime}} = 
\sum_{\kappa \kappa^{\prime}} 
U^{\text{nsf}}_{\mu \mu^{\prime} \kappa \kappa^{\prime}} [\hat{\chi}_{0}]_{\kappa \kappa^{\prime} \nu \nu^{\prime}} 
\end{equation} 
for $ \hat{V}^{\text{G.oB}}(q) $, 
and 
\begin{equation}
[ \hat{U} \hat{\chi}_{0} ]^{\text{nsf}}_{\mu \mu^{\prime} \nu \nu^{\prime}} = 
\sum_{\kappa \kappa^{\prime}} 
U^{\text{nsf}}_{\mu \kappa^{\prime} \kappa \mu^{\prime}} [\hat{\chi}_{0}]_{\kappa \kappa^{\prime} \nu \nu^{\prime}} 
\end{equation} 
for $ \hat{V}^{\text{G.L}}(q) $. 
For the non-spin-flip part with the 
on-site Hubbard interaction $V_{\alpha \alpha \alpha \alpha}$ 
and the interlayer Cooper pair hopping terms $V_{\alpha \beta \beta \alpha} \left( \alpha \neq \beta \right)$, 
the tensor products above are equivalent, and 
we arrive at
\begin{equation}
V^{G\text{.nsf}}_{\alpha^{\prime} \alpha \beta^{\prime} \beta}(q) 
= \Bigg[ \hat{U} +\frac{3}{2}\frac{\hat{U} \hat{\chi}_{0} \hat{U}}{1 -\hat{U} \hat{\chi}_{0}} 
                 +\frac{1}{2}\frac{\hat{U} \hat{\chi}_{0} \hat{U}}{1 +\hat{U} \hat{\chi}_{0}} 
                 -\hat{U} \hat{\chi}_{0} \hat{U} \Bigg]^{\text{nsf}}_{\alpha^{\prime} \alpha \beta^{\prime} \beta}(q), 
\end{equation}
where the second (third) term on the right-hand side 
is the spin- (charge-) fluctuation part. 

The anomalous self-energy for the interlayer interactions is given as 
\begin{equation}
-\Sigma^{F\text{.nsf}}_{\alpha \beta}(k) 
= \frac{1}{N\beta} \sum_{k^{\prime}} \sum_{\alpha^{\prime} \beta^{\prime}} 
  V^{F\text{.nsf}}_{\alpha^{\prime} \alpha \beta \beta^{\prime}}(k-k^{\prime}) 
  F_{\alpha^{\prime} \beta^{\prime}}(k^{\prime}),
\end{equation}
where 
\begin{align}
\hat{V}^{F\text{.nsf}} 
&= \hat{U}^{\text{nsf}} +\hat{V}^{F\text{.eB}} +\hat{V}^{F\text{.L}},  \\
V^{F\text{.eB}}_{\alpha^{\prime} \alpha \beta \beta^{\prime}}(q) 
&= \Bigg[ \frac{\hat{U} \hat{\chi}_{0} \hat{U} \hat{\chi}_{0} \hat{U}}
               {1 -\hat{U} \hat{\chi}_{0} \hat{U} \hat{\chi}_{0}} 
   \Bigg]^{\text{nsf}}_{\alpha^{\prime} \alpha \beta \beta^{\prime}}(q),  \\
V^{F\text{.L}}_{\alpha^{\prime} \alpha \beta \beta^{\prime}}(q) 
&= \Bigg[ \frac{\hat{U} \hat{\chi}_{0} \hat{U}}
               {1 -\hat{U} \hat{\chi}_{0}} 
   \Bigg]^{\text{nsf}}_{\alpha^{\prime} \alpha \beta \beta^{\prime}}(q), 
\end{align}
with the same rule for the tensor  products for $ \hat{V}^{\text{F.eB}}(q) $ and $ \hat{V}^{\text{F.L}}(q) $ 
as in the normal self-energy above. 
Therefore $ \hat{V}^{\text{F}}(q) $ is written as 
\begin{equation}
V^{F\text{.nsf}}_{\alpha^{\prime} \alpha \beta \beta^{\prime}}(q) 
= \Bigg[ \hat{U} +\frac{3}{2}\frac{\hat{U} \hat{\chi}_{0} \hat{U}}{1 -\hat{U} \hat{\chi}_{0}} 
                 -\frac{1}{2}\frac{\hat{U} \hat{\chi}_{0} \hat{U}}{1 +\hat{U} \hat{\chi}_{0}} 
  \Bigg]^{\text{nsf}}_{\alpha^{\prime} \alpha \beta \beta^{\prime}}(q). 
\end{equation}

\subsection{FLEX for Spin-flip interactions} 
Now we turn to the multiorbital FLEX with intra- and interlayer interactions belonging to the spin-flip $H_{\text{sf}}$. 
The normal self-energy for the interlayer interactions is given as
\begin{equation}
\Sigma^{G\text{.sf}}_{\alpha \beta}(k) 
= \frac{1}{N\beta} \sum_{k^{\prime}} \sum_{\alpha^{\prime} \beta^{\prime}} 
  V^{G\text{.sf}}_{\alpha^{\prime} \alpha \beta^{\prime} \beta}(k-k^{\prime}) 
  G_{\alpha^{\prime} \beta^{\prime}}(k^{\prime}).
\end{equation} 
For the spin-flip $H_{\text{sf}}$ 
we have to take account of all of bubble diagrams and odd numbers of ladders due to the spin selection rule in $H_{\text{sf}}$. 

However, we end up with the same form 
for the effective interaction for the self-energy $ \hat{V}^{G\text{.sf}}(q) $ as before, with separated spin and charge fluctuation parts, as 
\begin{equation}
V^{G\text{.sf}}_{\alpha^{\prime} \alpha \beta^{\prime} \beta}(q) 
= \Bigg[ \hat{U} +\frac{3}{2}\frac{\hat{U} \hat{\chi}_{0} \hat{U}}{1 -\hat{U} \hat{\chi}_{0}} 
                 +\frac{1}{2}\frac{\hat{U} \hat{\chi}_{0} \hat{U}}{1 +\hat{U} \hat{\chi}_{0}} 
                 -\hat{U} \hat{\chi}_{0} \hat{U} 
  \Bigg]^{\text{sf}}_{\alpha^{\prime} \alpha \beta^{\prime} \beta}(q). 
\end{equation} 
Similarly, 
the anomalous self-energy for the interlayer interactions is given 
as 
\begin{equation}
-\Sigma^{F\text{.sf}}_{\alpha \beta}(k) 
= \frac{1}{N\beta} \sum_{k^{\prime}} \sum_{\alpha^{\prime} \beta^{\prime}} 
  V^{F\text{.sf}}_{\alpha^{\prime} \alpha \beta \beta^{\prime}}(k-k^{\prime}) 
  F_{\alpha^{\prime} \beta^{\prime}}(k^{\prime}),
\end{equation}
where we have to take account of all of bubble diagrams and the even number of ladders 
due to the spin selection rule for $H_{\text{sf}}$. 

Thus we again end up with the same form for 
the pairing interaction for the anomalous self-energy as 
\begin{equation}
V^{F\text{.sf}}_{\alpha^{\prime} \alpha \beta \beta^{\prime}}(q) 
= \Bigg[ \hat{U} +\frac{3}{2}\frac{\hat{U} \hat{\chi}_{0} \hat{U}}{1 -\hat{U} \hat{\chi}_{0}} 
                 -\frac{1}{2}\frac{\hat{U} \hat{\chi}_{0} \hat{U}}{1 +\hat{U} \hat{\chi}_{0}} 
  \Bigg]^{\text{sf}}_{\alpha^{\prime} \alpha \beta \beta^{\prime}}(q), 
\end{equation}
with the spin- and charge-fluctuation parts. 

\subsection{Eliashberg equation} 
Finally, 
the normal and anomalous self-energies are written as 
\begin{equation} 
\begin{split} 
\Sigma^{G}_{\alpha \beta}(k) 
&= \frac{1}{N\beta} \sum_{k^{\prime}} \sum_{\alpha^{\prime} \beta^{\prime}} 
   V^{\text{eff}}_{\alpha^{\prime} \alpha \beta^{\prime} \beta}(k-k^{\prime}) 
   G_{\alpha^{\prime} \beta^{\prime}}(k^{\prime}),  \\ 
-\Sigma^{F}_{\alpha \beta}(k) 
&= \frac{1}{N\beta} \sum_{k^{\prime}} \sum_{\alpha^{\prime} \beta^{\prime}} 
   V^{\text{pair}}_{\alpha^{\prime} \alpha \beta \beta^{\prime}}(k-k^{\prime}) 
   F_{\alpha^{\prime} \beta^{\prime}}(k^{\prime}), 
\end{split} 
\end{equation} 
where $\hat{V}^{\text{eff}}$ and $\hat{V}^{\text{pair}}$ are expressed as Eq. (\ref{eq:tVeff}) and (\ref{eq:tVpair}), respectively. 
If we plug these into Dyson's equations for the anomalous Green's functions, 
we have the Eliashberg equation, 
\begin{equation} 
\begin{split}
\lambda \Delta_{\alpha \beta}(k) 
&= -\frac{1}{N\beta} \sum_{k^{\prime}} \sum_{\alpha^{\prime} \beta^{\prime}} \sum_{\gamma \delta} 
    V^{\text{pair}}_{\alpha^{\prime} \alpha \beta \beta^{\prime}}(k-k^{\prime})  \\ 
& \times G_{\alpha^{\prime} \gamma}(k^{\prime}) \Delta_{\gamma \delta}(k^{\prime}) G_{\beta^{\prime} \delta}(-k^{\prime}), 
\end{split} 
\end{equation} 
where $\hat{\Delta}(k)= \hat{\Sigma}^{F}(k)$.

\bibliographystyle{apsrev4-1} 

\end{document}